\documentclass[12pt]{iopart}
\usepackage{amssymb,bm}
\usepackage{graphicx}
\usepackage{dcolumn}
\usepackage{ifthen}
\usepackage{babel}
\usepackage{color}
\usepackage{caption}
\usepackage{float}
\usepackage{cite}
\usepackage{hyperref}
\hypersetup{
    colorlinks,
    citecolor=blue,    filecolor=blue,
    linkcolor=blue,     urlcolor=blue
}

\begin{document}

\title[Radiative characterization of supersonic jets and shocks]{Radiative characterization of supersonic jets and shocks in a laser-plasma experiment} 
\author{H Bohlin$^1$, F-E Brack$^{2,3}$, M Cervenak$^4$, T Chodukowski$^6$, J Cikhardt$^{4,7}$, J Dost\'{a}l$^{4,5}$, R Dud\v{z}\'{a}k$^{4,5}$, J. Hubner$^4$, W Huo$^{1,8}$, S Jelinek$^4$, D Kl\'{i}r$^{4,7}$, F Kroll$^2$, M Krupka$^{4,5}$, M Kr\r{u}s$^4$, T Pisarczyk$^6$, Z Rusiniak$^6$, U. Schramm$^{2,3}$, T-H Nguyen-Bui$^9$, S Weber$^{1,10}$, A Zara\'s-Szyd\l{}owska$^6$, K Zeil$^2$, D Kumar$^1$, T Schlegel$^1$, and V Tikhonchuk$^{1,9}$}

\address{$^1$ ELI-Beamlines, Institute of Physics, Czech Academy of Sciences,  25241 Doln\'{i} B\v{r}e\v{z}any, Czech Republic}
\address{$^2$ Helmholtz-Zentrum Dresden – Rossendorf, 01328 Dresden, Germany}
\address{$^3$ Technische Universität Dresden, 01062 Dresden, Germany}
\address{$^4$ Institute of Plasma Physics, Czech Academy of Sciences, 18200 Prague, Czech Republic}
\address{$^5$ Institute of Physics, Czech Academy of Sciences, 182 21 Prague, Czech Republic}
\address{$^6$ Institute of Plasma Physics and Laser Microfusion, 01-497 Warsaw, Poland}
\address{$^7$ Czech Technical University in Prague, 16627 Prague, Czech Republic}
\address{$^8$ Institute of Applied Physics and Computational Mathematics, 100094 Beijing, China}
\address{$^9$ Centre Lasers Intenses et Applications, University of Bordeaux -- CNRS -- CEA, 33405 Talence, France}
\address{$^{10}$ School of Science, Xi'an Jiaotong University, 710049 Xi'an, China}

\ead{hannes.bohlin@eli-beams.eu}

\vspace{10pt}
\begin{indented}
\item[]November 2020
\end{indented}

\begin{abstract}
The interaction of supersonic laser-generated plasma jets with a secondary gas target was studied experimentally. The plasma parameters of the jet, and the resulting shock, were characterized using a combination of multi-frame interferometry/shadowgraphy, and X-ray diagnostics, allowing for a detailed study of their structure and evolution. The velocity was obtained with an X-ray streak camera, and filtered X-ray pinhole imaging was used to infer the electron temperature of the jet and shock. The topology of the ambient plasma density was found to have a significant effect on the jet and shock formation, as well as on their radiation characteristics. The experimental results were compared with radiation hydrodynamic simulations, thereby providing further insights into the underlying physical processes of the jet and shock formation and evolution.   
\end{abstract}

\maketitle 
\section{Introduction}\label{sec1}
Supersonic jets and shocks are ubiquitous in astrophysics, and can, for example, be observed in accretion discs, Herbig-Haro objects, and the interaction of the solar wind with planets~\cite{McCray1979, McKee1980, Hartigan1989, Reipurth2001, Treumann2009}. While such phenomena cannot be directly reproduced in the laboratory, dedicated experiments allow for detailed studies of the fundamental mechanisms of their formation and evolution. By comparison of relevant dimensionless parameters such as the Mach and Euler numbers, the cooling parameter, and the jet-to-ambient plasma density ratio, the laboratory observations can be scaled to astrophysical phenomena that exhibit similar hydrodynamic behavior~\cite{Remington1999, Remington2006, Ryutov1999}. Laboratory experiments also provide benchmarks for the testing and development of multidimensional radiation hydrodynamic codes. 

A number of different approaches have been used for the experimental study of plasma jets, including high power lasers~\cite{Farley1999, Blue2005, Fox2013, Niemann2014, Hintington2015, Schaeffer2017}, and high-current Z-pinch facilities~\cite{Lebedev2002, Lebedev2019}. Generating jets using a laser and low atomic number targets, typically requires high-energy pulses with precise alignment over a conical target surface~\cite{Li2013}, or multi-compound target configurations~\cite{Foster2005}. By contrast, the iodine laser at the Prague Asterix Laser System (PALS) facility~\cite{PALS} allows for a relatively straightforward method of creating supersonic jets using a defocused laser on a target with high atomic number~\cite{Kasperczuk2006}. The jets are formed by the collision of ablated plasma converging on axis, which is due to a combination of the annular laser intensity profile on the target surface, and radiative cooling~\cite{ Kasperczuk2009}. The properties of these collimated supersonic plasma jets have been studied in several experiments~\cite{Nicolai2006, Kasperczuk2006, Pisarczyk2007, Kasperczuk2007,  Kasperczuk2007c, Kasperczuk2009,  Kasperczuk2009c}. It has also been shown that these jets can produce strong shocks when interacting with the plasma of a secondary gas target~\cite{Nicolai2008, Tikhonchuk2008, Nicolai2009, Kasperczuk2009b}.

Building on these previous experiments, an investigation was carried out on the formation and evolution of supersonic shocks. Using an extended set of optical and X-ray diagnostics, the jet-plasma interaction process was studied in detail, with particular emphasis on the radiative properties of the shock structure, and providing measurements of the electron temperature. The experimental observations are compared with radiation hydrodynamic numerical simulations, thereby providing further insights into the physical processes involved, as well as a validation of the code.

The experiment setup and diagnostics are outlined in Sec.~\ref{ExSetup}. The experimental results are described in Sec.~\ref{secResults}, starting with the structure and evolution of the jet and shock, followed by the characterization of the flow velocity and temperature using X-ray diagnostics. Numerical simulations of the jet and shock are presented in Sec.~\ref{Sim_sec}, and compared with the experimental observations. Lastly, Sec.~\ref{SumSec} presents a short summary and conclusions, and the scaling of the experiment to astrophysical phenomena.

\section{Experiment setup}\label{ExSetup}

\subsection{Laser and target configuration}\label{ExSetup1}
The experiment was carried out with the iodine laser at PALS~\cite{PALS}, with the third harmonic of the beam ($\lambda = 0.438\,\mu$m) used for generating the plasma jets. The laser pulse had a Gaussian temporal profile with a full width at half maximum (FWHM) of 250\,ps. For these experiments, either the full aperture beam diameter of 290\,mm, or an apodized diameter of 190\,mm, was used. It was focused using an aspheric lens ($f = 600$\,mm) at normal incidence onto a flat massive copper target, with the focal plane located slightly behind the front surface of the target. The resulting defocused irradiated area on the target had a radius of $r_{\rm L}\simeq 300\,\mu$m, with an annular intensity distribution~\cite{Kasperczuk2009, Kmetik2012}. The laser energy was in the range of $E_{\rm L}\simeq 70 - 110$\,J, giving an average intensity on target of $I_{\rm L} \lesssim 10^{14}$\,Wcm$^{-2}$. Similar focusing conditions have been used in previous experiments at PALS, and the interaction parameters were optimized for jet formation~\cite{Nicolai2006, Pisarczyk2007, Nicolai2008}. 

\begin{figure}[b!]
\centering
\includegraphics[width=0.95\linewidth]{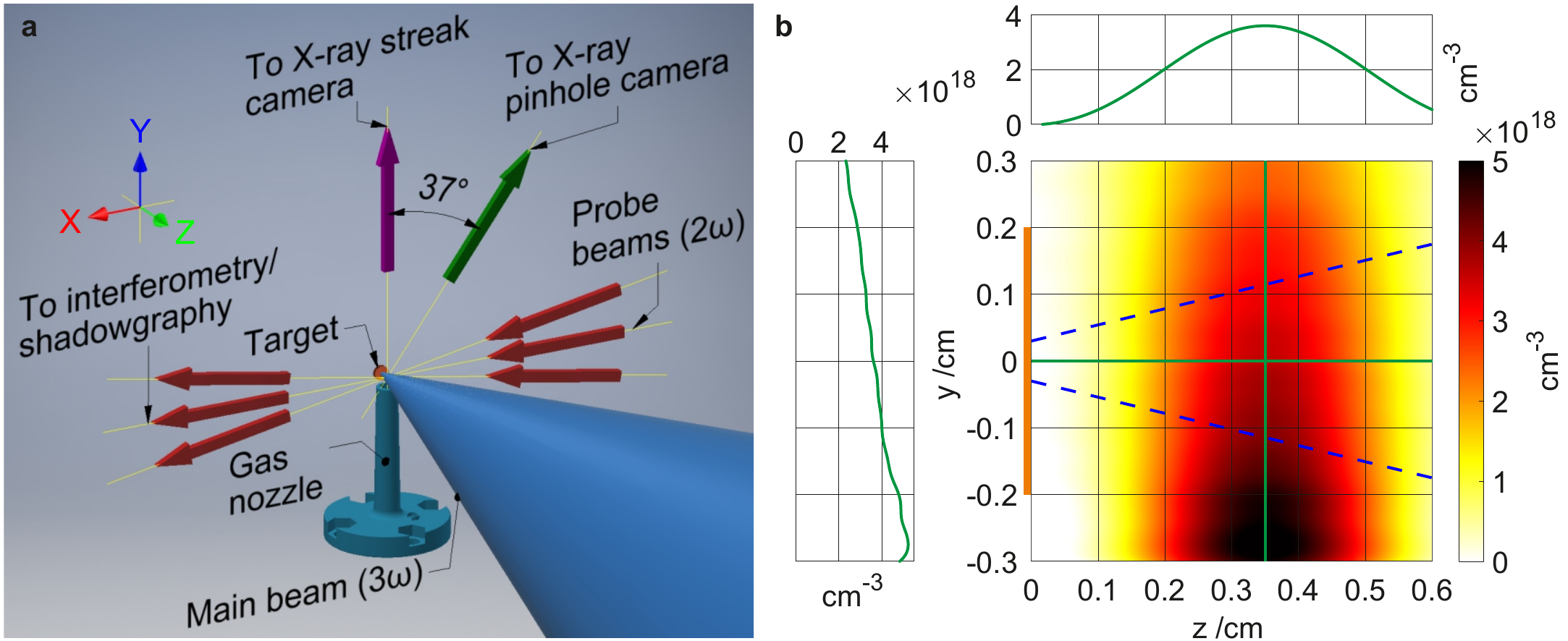}
\caption{a) Experiment setup showing the copper target, the gas nozzle, the main laser beam, and the line-of-sights of the diagnostics. b) The Abel inverted 2D map of the neutral density distribution of the gas target measured with interferometry using a CW He-Ne laser. Also shown are line-outs of the density along the target normal at $y = 0$, and along the $y$-axis at the center of the nozzle orifice ($z = 3.5$\,mm). The target location is given by the orange line at $z = 0$, and the blue dashed lines indicate the laser cone.}	\label{Setup}
\end{figure}

The setup of the experiment is shown in figure~\ref{Setup}a. The shocks were formed through the collisional interaction of the plasma jet with a secondary argon gas target produced with a de Laval nozzle. The neutral density profile of the nozzle, measured with interferometry using a continuous-wave (CW) Helium-Neon (He-Ne) laser, is shown in figure~\ref{Setup}b. The geometry of the laser-target setup is also shown. The origin is taken as the center of the front surface of the copper target, as indicated by the orange line, and the blue dashed lines indicate the edges of the laser cone. The orifice of the nozzle had a diameter of 3.8\,mm, with its center at $z = 3.5$\,mm and $y = -3$\,mm. 

A backing pressure in the range of $p = 5 - 40$\, bar was used, and the neutral density, which scales linearly with pressure, was on the order of $n_{\rm Ar}\sim 4 \times 10^{18}$~cm$^{-3}$ in the interaction region for $p =\,40$\,bar (see line-outs in figure~\ref{Setup}b).

\subsection{Diagnostics}\label{ExSetup2}

The jet and shock parameters were characterized using a combination of optical and X-ray diagnostics, consisting of a three-frame optical probing system used either for shadowgraphy and/or interferometry, a four-frame X-ray pinhole camera, and an X-ray streak camera. The lines of sight of the diagnostics are shown in figure~\ref{Setup}a.

The optical imaging diagnostic utilized the second harmonic of the laser split into three probe beams, with adjustable delays between the three frames, and a temporal resolution of 250\,ps. The diagnostic had a line-of-sight perpendicular to the target normal, and a field-of-view of 6\,mm. Interferometry was used to obtain the line-integrated electron density, and shadowgraphy was used to study the structure of the shock in greater detail, as it provided the better spatial resolution of the two. The electron density range accessible for the optical diagnostics was approximately $n_e \sim 10^{17} - 10^{20}$\,cm$^{-3}$. Due to the up-down asymmetry of the jets and the shocks observed in the experiment, Abel inversion was not implemented for obtaining their two-dimensional electron density profile. However, the density along the jet and shock axis could be estimated from the average of the Abel inversion performed on the top and bottom half of the line-integrated measurements separately. The error, based on the typical difference in the on-axis value between the two halves, was approximately 20\%.

The X-ray pinhole camera had a line-of-sight perpendicular to the target normal, and an angle of $37^\circ$ with respect to the vertical axis in the target plane, as shown in figure~\ref{Setup}a. It utilized four elliptical pinholes, with one covered with a 0.2\,$\mu$m thick iron foil. The pinhole images were projected onto a gold-coated microchannel plate (MCP) detector with four quadrants, which was coupled to a phosphor screen imaged with a Nikon D1X CCD digital camera. The exposure time was 2\,ns FWHM, and the adjustable delays between quadrants set up to record at three different time instances, with two of the frames simultaneously measuring the open and filtered emission. The MCP is sensitive in the photon energy range of $E_{\rm ph} = 10 - 1000$\,eV, and while the exact spectral response is not known, it can be approximated by the quantum efficiency (QE) of a gold photocathode. For this purpose, the QE measurements by Henneken et al. have been used~\cite{Henneken}. 

The X-ray streak camera had a top view, with the input slit oriented along the target normal, providing a field of view of $\Delta z \sim 3.5$\,mm, and a streak time duration of $\Delta t = 18$\,ns. 

\section{Experimental results}\label{secResults}

\subsection{Structure of the plasma jet}\label{secjet}

\begin{figure}[!t]
\centering
\includegraphics[width=0.95\linewidth]{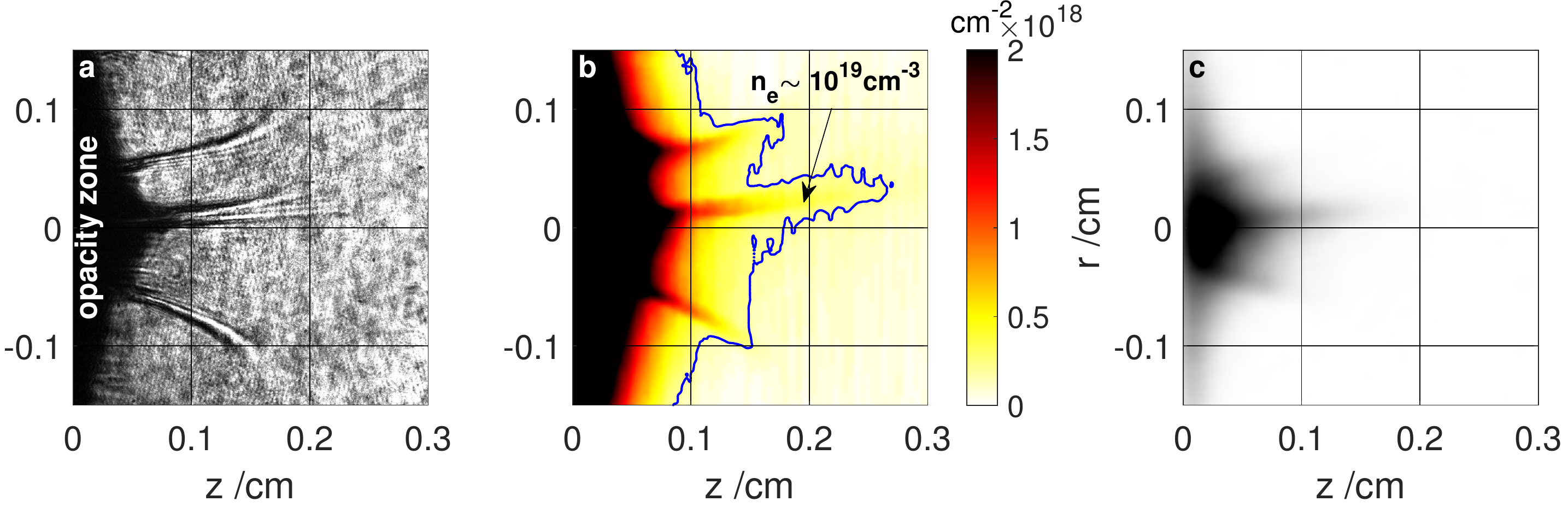}
\caption{Plasma jet created by the interaction of the laser with a copper target. a) Shadowgraphy, where the darker regions correspond to locations with high electron density gradients; and b) the simultaneously measured line-integrated electron density ($t = 6.5$\,ns, and $E_{\rm L}= 71$\,J). The blue contour line indicates $N_e = 2 \times 10^{17}$\,cm$^{-2}$. The electron density is approximately $n_e \sim 10^{19}$\,cm$^{-3}$ at the location indicated by the arrow. c) Unfiltered X-ray pinhole image of a jet at $t = 3$\,ns, for a shot with $E_{\rm L} = 82$\,J.} 
\label{Jet}
\end{figure}

Shadowgraphy of a typical jet is shown in figure~\ref{Jet}, along with the simultaneously measured line-integrated density, $N_e=\int n_e dl$, obtained with interferometry ($t= 6.5$\,ns, and $E_{\rm L}= 71$\,J). An example of the X-ray emission recorded with the pinhole camera is also shown ($t = 3$\,ns, and $E_{\rm L} = 82$\,J). The laser is incident from the right in the images, and the target front surface is located at $z = 0$. The main features of the jet consist of a high-density region of ablated copper near the target, which is opaque to the optical diagnostics, and a central collimated plasma flow extending along the target normal at a slight angle. The deviation of the jet is likely from an uneven radial plasma convergence on axis due to an asymmetry of the annular laser intensity distribution on target. Two smaller jet-like structures can also be observed on each side of the central jet. These likely originate from the collision of the radially expanding hot central plasma with a colder plasma produced at the edges of the laser spot. The electron density of the jet in figure~\ref{Jet}b is on the order of $n_e\simeq 10^{19}$\,cm$^{-3}$, with a radius of $r_J \sim 0.1 - 0.15$\,mm. At this time, the jet extends to approximately $z \sim 3$\,mm, beyond which the density is below the detection limit of the diagnostic. The jet velocities observed in the experiment were in the range of $v_{\rm J}\sim 400-700$\,km\,s$^{-1}$, with a lifetime of more than $10 - 15$\,ns after the end of the laser pulse, depending on the irradiation conditions.

\newpage
\subsection{Spatial and temporal evolution of the shock structure}\label{secResults2}

\subsubsection{Characterization of the shock structure with the optical probing system}\label{secResults21}\hfill\\

\begin{figure}[!t]
\centering
\includegraphics[width=0.95\linewidth]{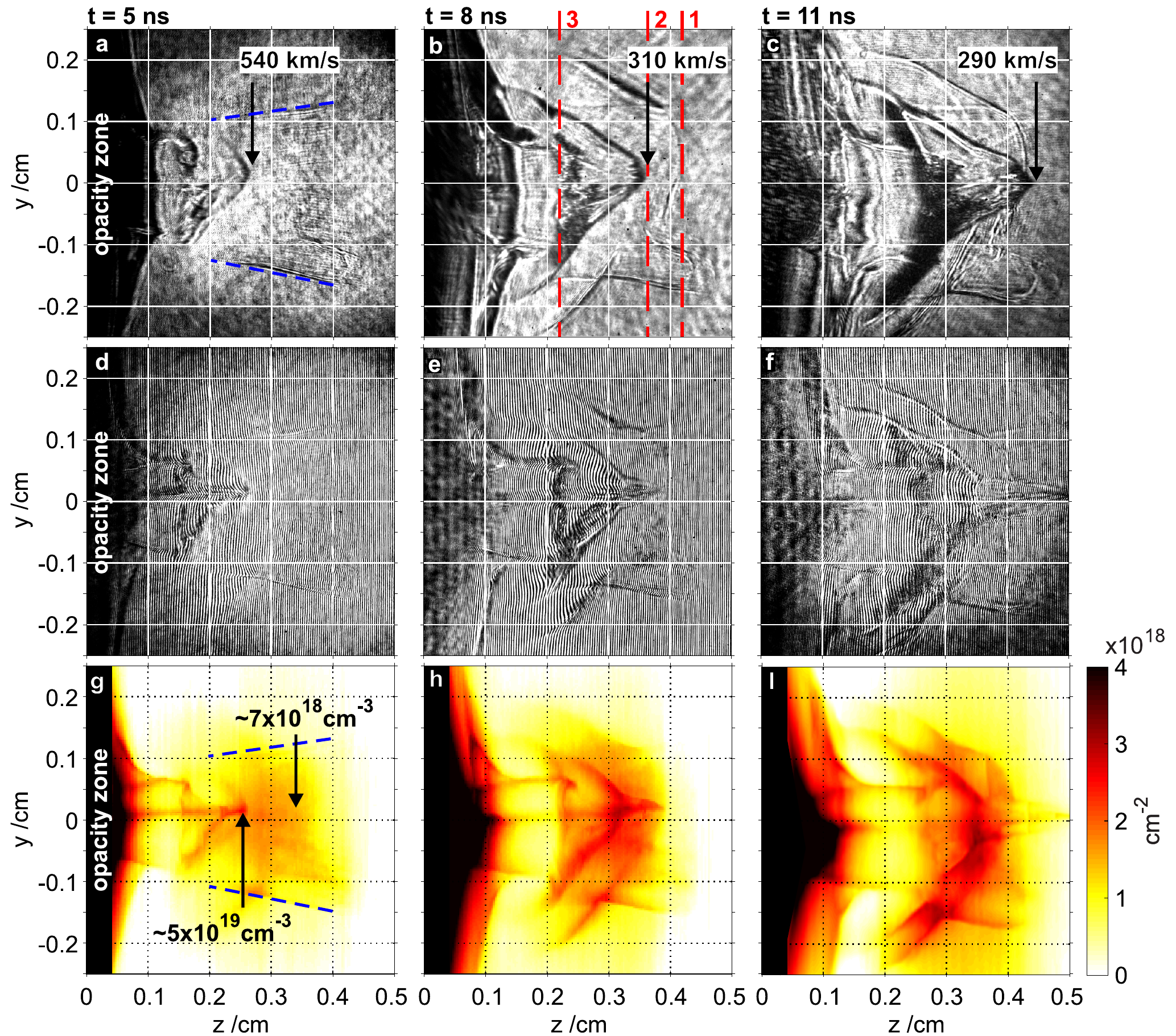}
\caption{Shocks formed through the interaction of the copper jet with the argon plasma. Shadowgraphy (top), and line-integrated electron density measured with interferometry (bottom) at $t = 5$, 8 and 11\,ns ($E_{\rm L} = 110$\,J and $E_{\rm L} = 108$\,J, respectively). The raw interferograms are included for reference (middle). A pressure of $p = 20$\,bar was used for both shots. The blue lines in figure a and g show the boundaries of the region of the laser induced ionization. The red lines in figure b indicate the bow shock (1), and the boundaries of the reverse shock (2 and 3). The velocity of the copper plasma at the point indicated by the arrows, is also given in the shadowgraphy images.}\label{Shock}
\end{figure}

The jets were used to generate shocks through the interaction with the secondary argon gas target. Shadowgraphy, and the line-integrated electron density, measured at three time instances, are shown in figure~\ref{Shock} for shocks created using a gas pressure of $p = 20$\,bar, and similar laser energies ($E_{\rm L} \sim 110$\,J). The raw interferograms are also included for reference. In the initial phase preceding the jet formation, the argon gas is partially ionized though a combination of thermal X-ray emission from the ablated copper plasma, and by subsequent collisional absorption from the laser traversing the gas. The laser heated region is bounded by the converging traces seen in the gas region, which are indicated by the dashed lines in figure~\ref{Shock}a and ~\ref{Shock}g. They are most prominent in the lower half, where the gas density is higher. The background argon electron density, which depends on the backing pressure and laser energy used, was approximately $n_e \sim 5 \times 10^{18}$\,cm$^{-3}$ in the central region for these type of shot parameters.

The structure of the interaction zone was characterized in previous experiments~\cite{Nicolai2008, Tikhonchuk2008}, and for consistency, the same definitions will be used here. The main shock features are indicated by the red dashed lines in figure~\ref{Shock}b, and consist of: (1) the bow shock that propagates in the argon plasma; (2) the leading edge of the copper plasma, which together with (3) give the boundaries of the reverse shock (Mach disk) that slows the jet and results in an accumulation of copper plasma downstream. The working surface (WS) of the shock corresponds to the region bounded by the bow shock and the Mach disk. The jet is enveloped in a cocoon of copper plasma flowing backwards from the working surface. As demonstrated in the simulations in section~\ref{Sim_sec}, the copper and argon plasma do not interpenetrate, but form a contact surface with a shear flow.

The shock structure is highly dependent on the jet topology, as well as the non-uniform argon density profile. The latter can be divided into the central laser heated region, and the two X-ray heated regions below and above. As a result, the shock has a location-dependent variation of the thickness and opening angle of the WS, with the narrowest part seen in the lower half, where the neutral gas density is higher. The background plasma is also less uniform inside than outside the laser heated region, and the density gradient of the argon shock is shallower here. As a result, the bow shock is harder to distinguish in this region in the shadowgraphy frames. From these observations, it is clear that the laser induced ionization, and its effect on the topology of the background plasma, influences the shock structure. Similar results were seen in the shock simulations presented in section~\ref{Sim_sec}. This observation is also supported by previous shock experiment at PALS~\cite{Kasperczuk2009b}, where minimizing the laser interaction with the gas target by using an oblique beam incidence, resulted in an improved uniformity of the background plasma, and a better resolved shock structure. 

Estimates of the electron density at $t = 5$\,ns are shown in figure~\ref{Shock}g at the points corresponding to the leading edge of the copper plasma, and the argon bow shock. The latter has a density of $n_e\sim 7\times 10^{18}$\,cm$^{-3}$, and assuming a background electron density of $n_e\sim 5\times 10^{18}$\,cm$^{-3}$, this implies a relatively weak shock compression ratio of 1.4 in the argon. In contrast, the density of $n_e\sim 5\times 10^{19}$\,cm$^{-3}$ at the leading edge of the copper plasma, is an order of magnitude greater than the initial argon background density at this location. It also corresponds to a factor of five increase compared to the jet density of $n_e\sim 10^{19}$\,cm$^{-3}$.

In the shock evolution shown in figure~\ref{Shock}, the argon bow shock appears to stagnate in the region of maximum neutral density, reaching a distance of approximately $z \sim 4$\,mm. By contrast, the jet, and accumulated copper plasma in the reverse shock, retain enough momentum to catch up to the argon bow shock at $t = 8$\,ns, and has broken through it at $t = 11$\,ns. In the process, the working surface of the shock narrows from $\Delta z > 1$\,mm at $ t= 5$\,ns, to $\Delta z < 0.5$\,mm at $t= 11$\,ns. The compression in argon also increases, but is less straightforward to evaluate due to the jet overtaking the argon bow shock.

\subsubsection{Characterization of the shock structure with the X-ray pinhole camera}\label{secResults22}
\begin{figure}[!b]
\centering
\includegraphics[width=0.95\linewidth]{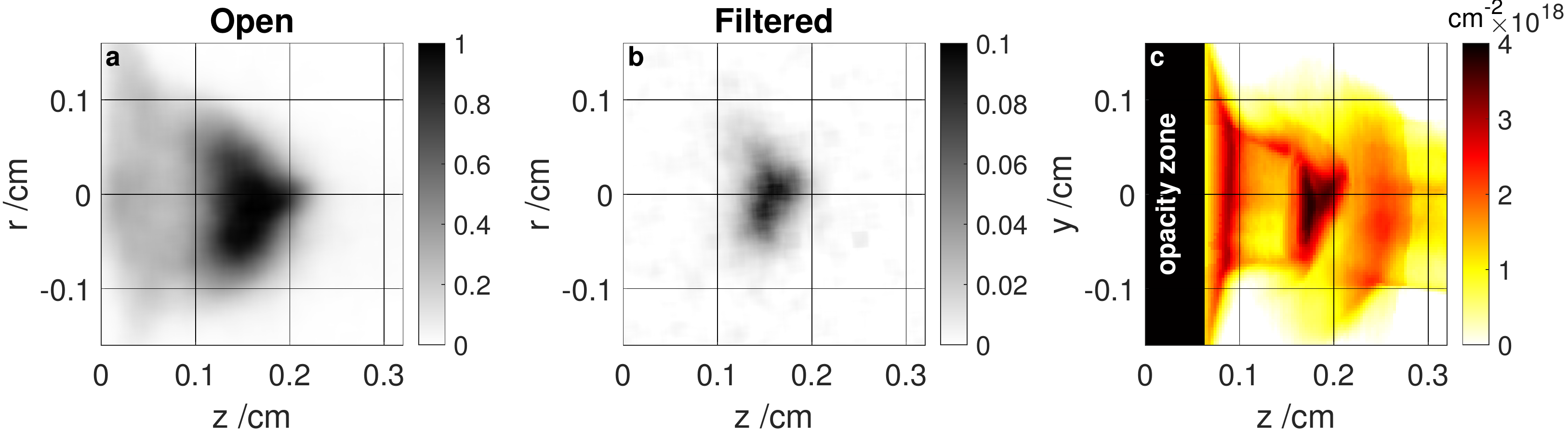}
\caption{a) Open, and b) filtered images taken with the X-ray pinhole camera at $t = 5$\,ns. c) The simultaneously measured line-integrated electron density. An argon gas pressure of $p = 40$\,bar, and laser energy $E_{\rm L} = 85$\,J (with apodized beam) was used for this shot.}
\label{PinholeShock}
\end{figure}

\hfill\\
The open and filtered X-ray emission of a shock imaged with the X-ray pinhole camera, together with the simultaneously measured line-integrated density, are shown in figure~\ref{PinholeShock}. A pressure of $p = 40$\,bar, and laser energy of $E_{\rm L} = 85$\,J, were used for this shot, and the images were taken at $t = 5\,$ns. The filtered imaging allows for the regions with the hottest emission to be differentiated. For this pressure and time, the shocked zone appears brightest, with the target and jet barely distinguishable in the unfiltered frame, and not seen at all in the filtered one. For both frames, the strongest emitting region extends to a maximum distance of $z \sim 2.1$\,mm from the target. Comparing with the interferometry in figure~\ref{PinholeShock}c, the filtered emission has similar dimensions and location as the shocked copper. It can therefore be concluded that the emission primarily originates from the copper plasma, with little argon emission seen in the open frame in comparison, and none at all in the filtered one.

\begin{figure}[!t]
\centering
\includegraphics[width=0.95\linewidth]{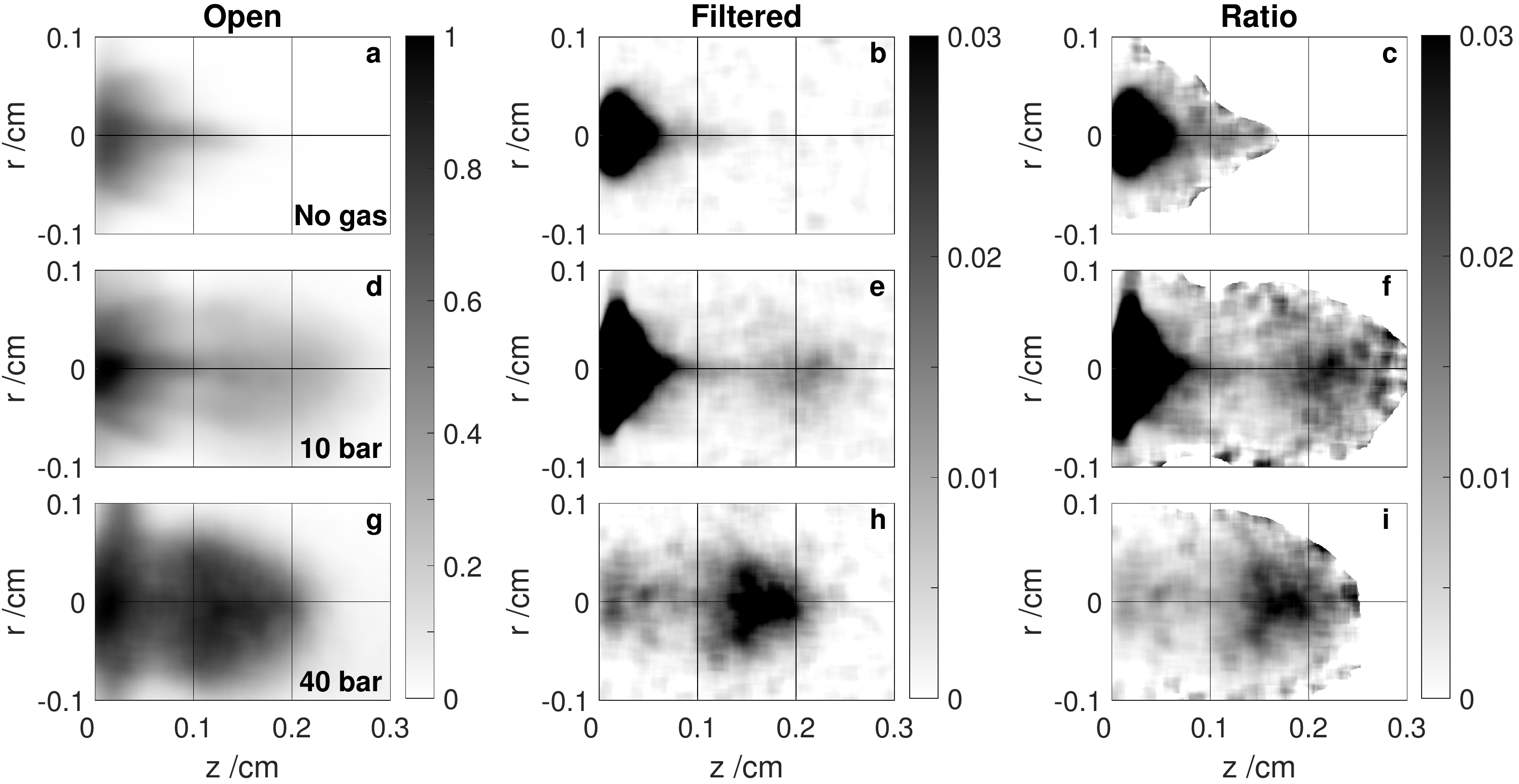}
\caption{Open (left column), and filtered X-ray pinhole images (central column), and their ratio (right column) at $t=3$\,ns for different argon pressures and similar laser energies. The top row shows the jet without the gas ($E_{\rm L} = 104$J), the middle row for $p = 10$\,bar ($E_{\rm L} = 109$\,J), and the bottom row for $p =  40$\,bar ($E_{\rm L} = 104$\,J). The data is normalized to the maximum in frame g. The images have been rotated to align the jet propagation with the $z$-axis. For the right column, the data outside the regions of interest has been omitted due to the low and noisy signal there.}	\label{Ratio2D}
\end{figure}

To further study the influence of the gas pressure on the shock structure observed with the X-ray pinhole camera, measurements at similar laser energies and times ($t = 3\,$ns) were done for backing pressures of $p = 10$\,bar, and $p =  40$\,bar. This is shown in figure~\ref{Ratio2D}, with the case of the jet without the gas in the top row for reference. A number of observations can be made from these images. To begin with, the filtered emission near the target is brighter for 10\,bar than for the other two cases. The jet also appears narrower and extends further from the target than without gas. By contrast, for the highest pressure of 40\,bar, the strongest emission is observed in the shocked material, which is also located closer to the target at this time. Furthermore, the filtered emission from the jet-forming region near the target has already decreased significantly after 3\,ns. In comparison, depending on the irradiation conditions, the duration of the emission at the target without a gas can exceed 10\,ns in the open frame and 5\,ns in the filtered one. This implies a relatively long cooling time in the experiment.  

These observations are in agreement with previous shock experiments at PALS, where shadowgraphy and interferometry showed a decrease in the working surface thickness, as well as the distance between the bow shock and target, for increasing gas pressure \cite{Nicolai2008}. Similar to the present experiment, an improved jet collimation was also observed at low gas pressure (e.g. 5\,bar). While it is clear from figure~\ref{Ratio2D} that the gas pressure affects not only the shock structure, but also the jet formation, the question remains as to what mechanisms are responsible for these observations. One explanation, based on the simulations discussed in section~\ref{Sim_sec}, is that the background argon plasma, and the shocked material flowing around the jet, restrict radial motion, and thereby redirect and collimate the flow of copper. Since the argon gas extends all the way to the target, it would be expected to influence the jet also in the initial stages, which would explain the reduced jet radius, and increase in filtered emission at the target, for 10\,bar. This suggests the possibility of using a low density gas near the target to facilitate the formation of narrow well-collimated jets. For a sufficiently high pressure, the smaller jets at the edge of the focal spot also become diverted, further affecting the jet and shock structure. For the high pressure case in figure~\ref{Ratio2D}, the axial motion is also significantly influenced by the gas, and hence the shock develops earlier and closer to the target.

\subsection{Characterization of the plasma flow velocity}\label{secResults3}

The velocity of the copper plasma can be evaluated by tracking the location of its leading edge between shadowgraphy frames in figure~\ref{Shock}. For the first frame, which corresponds to the distance crossed during the first 5\,ns after the laser pulse, a velocity of $v = 540\pm40$\,km\,s$^{-1}$ is obtained. Here, the error is given by the uncertainties in the position of the leading edge, and the registration time of the image. The density front of the copper then slows down as it traverses the gas target, and reaches an approximately constant velocity of $v = 300\pm70$\,km\,s$^{-1}$. This can be attributed to a combination of the interaction with the argon plasma, and the jet velocity plateauing as the target cools down~\cite{Kasperczuk2006}.

\begin{figure}[t!]
\centering
\includegraphics[width=0.95\linewidth]{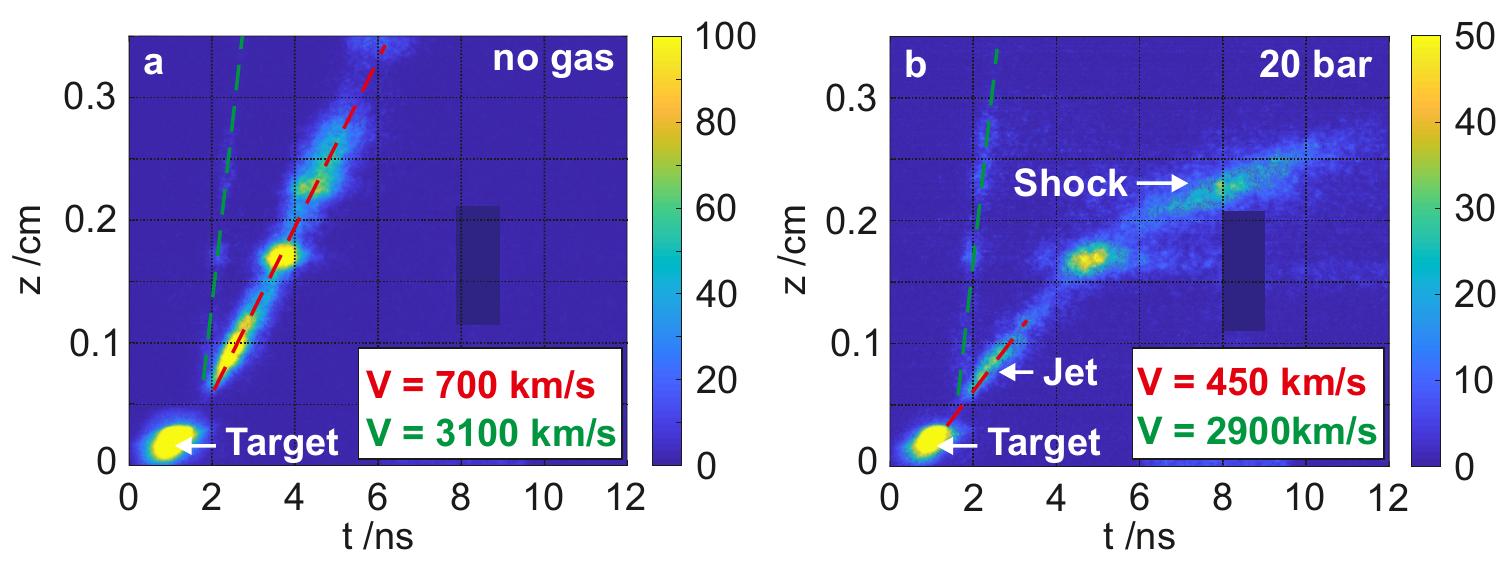}
\caption{X-ray streak camera images: a) without gas ($E_{\rm L} = 81$\,J), and b) at gas pressure of 20\,bar ($E_{\rm L} = 93$\,J). The dashed lines show linear fits along the maxima of the emission, which are used to obtain the velocity of the plasma flow.}\label{streak}
\end{figure}

The jet and shock were also imaged with the X-ray streak camera, and measurements without the gas target, and at a backing pressure of $p = 20$\,bar are shown in figure~\ref{streak}. The diagnostic exhibited a non-uniform flat field, and hence no quantitative estimates of the photon intensity could be derived from the streak camera measurements. However, the location of the emission from the ablated copper plasma can be clearly distinguished in each image, and the velocity can be estimated from the gradient of a line fitted along the maximum of the X-ray streak. The measured jet velocity is then $v_{\rm J}= 700\pm90$\,km\,s$^{-1}$, where the error is estimated from the spread of the streak signal around the peak value. This is consistent with the velocity obtained from interferometry/shadowgraphy measurements.

For the case with the gas shown in figure~\ref{streak}b, a reduced initial jet velocity of $v_{\rm J} = 450\pm50$\,km\,s$^{-1}$ was observed, which also agrees with the shadowgraphy in figure~\ref{Shock}. This could be a result of the low density gas near the target influencing the jet formation, and slowing it down as it propagates towards the high density region at the nozzle. When the copper plasma reaches the edge of the argon plasma, the velocity is reduced further. While the exact shock velocity is difficult to determine due to the spread of the signal and the location-dependent sensitivity of the diagnostic, the shock formation is clearly distinguished. Furthermore, as with the X-ray pinhole camera images, there is little emission seen beyond $z = 3$\,mm, which further supports the assumption that the observed emission is mainly from the copper plasma. 

Preceding the jet, X-ray emission propagating at a velocity of $v\sim 3000$\,km\,s$^{-1}$ is observed (green dashed line in figure~\ref{streak}). This is comparable to the expected electron thermal velocity of a laser-irradiated copper plasma, and could therefore be associated with a beam of electrons ejected from the target plasma during the laser pulse. However, this hypothesis requires further investigation to determine its validity.

\subsection{Measurements of plasma electron temperature with filtered X-ray imaging}\label{Temp_sec}

While the individual frames of the X-ray pinhole camera do not directly provide quantitative information of the photon intensity of the emission, the ratio of the unfiltered and filtered images gives the relative contribution in the photon energy range for which the filter is transmitting. An estimate of the temperature of the jet and the shock can then be obtained by comparing this ratio with the spectral intensity obtained from atomic models. This is in contrast to other radiative shock experiments, which rely on Thomson scattering~\cite{Seo2018}, or on radiative hydrodynamic simulations for estimating the electron temperature~\cite{Remington2006, Michaut2009}. This method of filtered Bremsstrahlung measurements have been used in magnetic fusion plasmas to infer sub keV electron temperatures~\cite{Delgado2007}. However, while the low density plasma in those experiments can be assumed to be in coronal equilibrium, a collisional-radiative model is better suited for the dense plasma in the present experiment.

\begin{figure}[t!]
\centering
\includegraphics[width=0.9\linewidth]{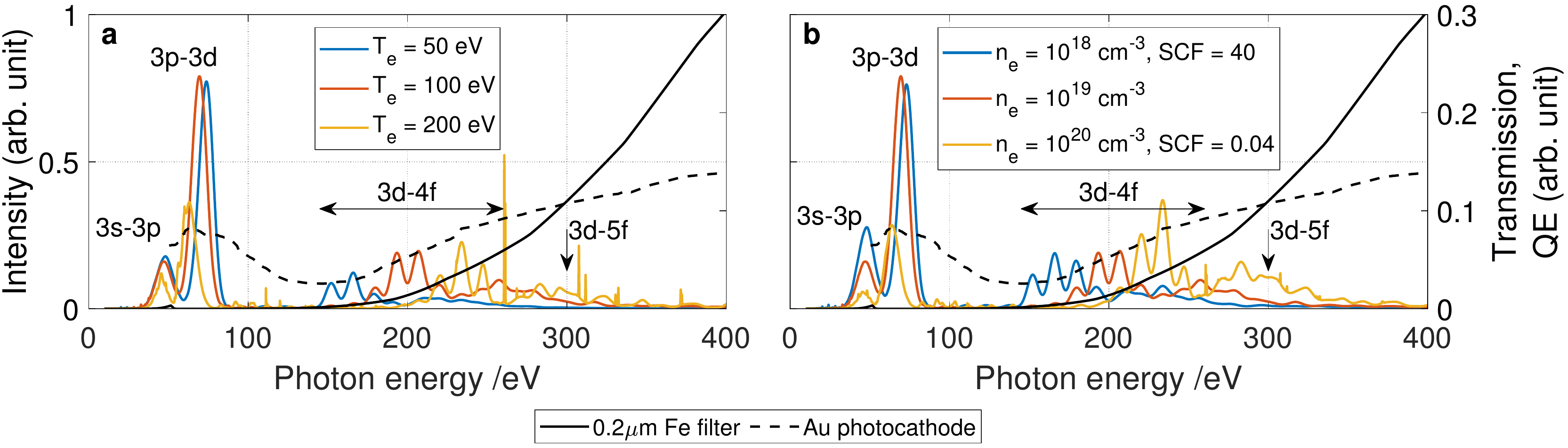}
  \caption{Spectral intensity of emission from Cu plasma simulated with FLYCHK for varying: a) electron temperatures ($n_e = 10^{19}$\,cm$^{-3}$); and b) electron densities ($T_e = 100$\,eV). For clarity of presentation, a scaling factor (SCF) has been used for some spectra. The solid black lines show the transmission through a 0.2\,$\mu$m thick iron foil \cite{Henke:LBL}, and the dashed black lines the quantum efficiency (QE) of a gold photocathode adapted from the measurements by Henneken et al.~\cite{Henneken}}.
  \label{Cu-spectra}
\end{figure}

The experimental data was interpreted using simulations performed with the atomic code FLYCHK~\cite{FLYCHK}, and the spectra were generated using the accompanying code FLYSPEC. Examples of simulated unfiltered spectra for a copper plasma at varying temperatures (for $n_e = 10^{19}$\,cm$^{-3}$), and densities (for $T_e = 100$\,eV), are shown in figure~\ref{Cu-spectra}a and \ref{Cu-spectra}b, respectively. The transmission curve of the 0.2\,$\mu$m thick iron filter is also shown for reference, along with the quantum efficiency (QE) for a gold photocathode adapted from measurements by Henneken et al.~\cite{Henneken}, which has been used to approximate the spectral response of the MCP. As expected, the spectrum shifts to higher photon energies for increasing electron temperatures. At $T_e = 50$\,eV the emission is mainly below $E_{ph} \sim 70$\,eV, and dominated by the $3s-3p$ and $3p-3d$ transitions. For higher temperatures, the electrons are energetic enough to excite the copper ions to the $n = 4$ and 5 levels, and the $3d-4f$, and $3d-5f$ transitions contribute significantly to the total intensity. Most of this emission is above $E_{\rm ph} \sim 170$\,eV, and therefore transmitted through the filter. Consequently, observation of filtered emission with the pinhole camera is direct evidence of higher electron temperatures, and the ratio with the unfiltered emission can be used to estimate the electron temperature of the plasma. As seen in figure~\ref{Cu-spectra}b, increasing the electron density also results in significantly higher electron excitation rates of the Cu ions, and consequently an overall increase in the transmission through the filter.

\begin{figure}[!t]
\centering
\includegraphics[width=0.95\linewidth]{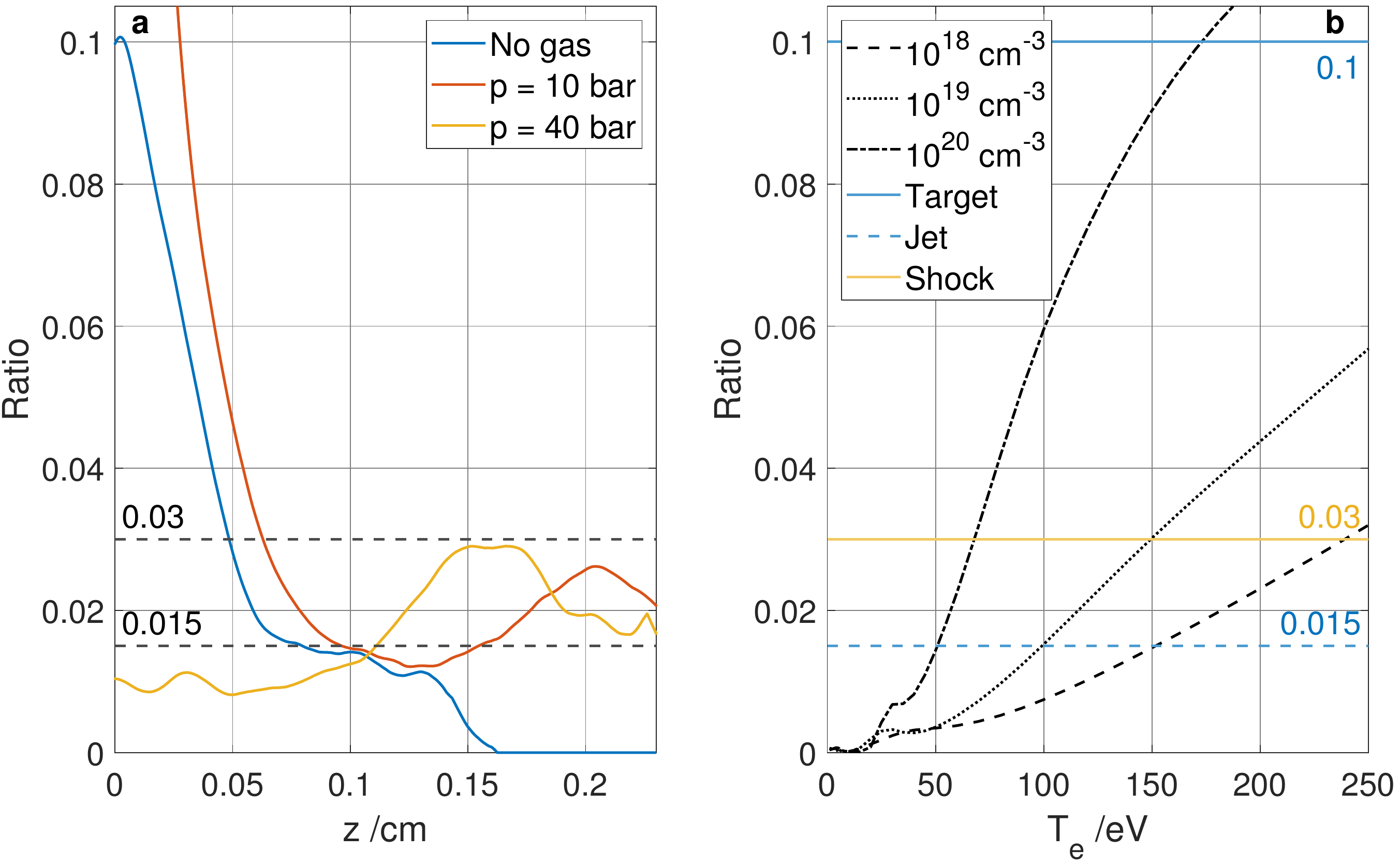} 
\caption{a) Line-outs of the ratios in figure~\ref{Ratio2D} along the jet axis. b) Dependence of the ratio of integrated FLYCHK spectra on copper plasma temperature for different electron densities ($n_e= 10^{18}-10^{20}\,$~cm$^{-3}$).}	\label{RatioComp}
\end{figure}

The simulated spectra were adjusted based on the QE of a gold photocathode~\cite{Henneken}, and integrated over the photon energy range $10 - 1000$\,eV for which the MCP is sensitive. The resulting temperature dependence of the ratio of filtered and unfiltered X-ray intensities, $I_{\rm f}/I_{\rm o}$, is given in figure~\ref{RatioComp}b. An estimate of the electron temperature can then be obtained by comparison of the measured emission ratio with the corresponding calculated value. 

The ratios of the previously described open and filtered X-ray pinhole frames (no gas, 10\,bar, and 40\,bar), are shown in figure \ref{Ratio2D}c, f, and j, and line-outs along the jet propagation are shown in figure \ref{RatioComp}a. For the case of the copper jet, a ratio of $I_{\rm f}/I_{\rm o} > 0.1$ is obtained at the target at $t = 3$\,ns. Comparing with the simulated ratio, this yields an electron temperature estimate of $T_e \sim 175$\,eV for $n_e = 10^{20}$\,cm$^{-3}$. Since the electron density can be assumed to be greater than that at the target, this can be taken as an upper bound of the temperature in this region. Away from the target, the ratio drops to $I_{\rm f}/I_{\rm o} = 0.015$, corresponding to a jet temperature in the range of $T_e = 50 - 150$\,eV, depending on the density ($T_e \sim 100$\,eV for $n_e= 10^{19}$\,cm$^{-3}$). 

For the cases with the gas target, the increase in the ratio at distances $z > 1$\,mm correspond to the shocked plasma. Since the emission in the shock can be from both copper and argon, with their relative contribution unknown, the temperature evaluation becomes less straightforward. However, as the brightest emission is located in the region of the reverse shock, the copper emission should dominate. Based on this assumption, the ratios $I_{\rm f}/I_{\rm o}\sim 0.027$ and $I_{\rm f}/I_{\rm o}\sim 0.03$ for argon pressures of 10\,bar and 40\,bar, respectively, gives a shocked copper electron temperature in the range of $T_e = 70 - 250$\,eV. The density in the shock usually exceeds $n_e = 10^{19}$\,cm$^{-3}$, and hence the upper bound of the electron temperature is likely on the order of $T_e \sim 150$\,eV.

Since the filtered emission of the background plasma is below the noise level of the diagnostic, the above described method could not be used for obtaining the temperature in this region. Nevertheless, the temperature of the background argon plasma can be inferred from the expected degree of ionization of the gas. As an example, the background electron density at $z \sim 3.5$\,mm in figure~\ref{PinholeShock}c, is on the order of $n_e \sim 7 \times 10^{18}$\,cm$^{-3}$. As the neutral gas density for a pressure of $p = 40$\,bar was around $n_{\rm Ar} = 4 \sim 10^{18}$\,cm$^{-3}$, the average degree of ionization in the background argon plasma can be assumed to have been approximately $Z \sim 1-2$. This corresponds to an electron temperature on the order of a few electronvolts at most.

The absence of emission from the shocked argon in the filtered frame of the X-ray pinhole camera, may similarly have been due to a low electron temperature. However, opacity calculations with FLYCHK suggest that the emission could have also been attenuated by re-absorption as the photons traversed the layer of colder ambient argon plasma in front of the detector. Consequently, no definitive conclusion can be drawn regarding the temperature in the argon shock based on these X-ray measurements.

\section{Simulations}\label{Sim_sec}

\subsection{Model description and input parameters}

The evolution of the supersonic plasma jet ejected from the copper target, and its interaction with the argon gas, were simulated using the radiation-hydrodynamic code FLASH~\cite{Fryxell_2000, FLASH1, FLASH2}. The code works with a fixed, finite-volume Eulerian grid, and uses arbitrary mesh refinement. In the simulations presented here, the code was used in a two-dimensional cylindrical-symmetric geometry, with a spatial resolution better than 10\,$\mu$m. Standard boundary conditions were applied for the solutions of the hydrodynamic equations, and were reflecting at the symmetry axis, and outgoing for the plasma flow, radiation, and thermal transport. The plasma is described in a three-temperature approximation due to possible local thermal non-equilibrium of electrons, ions and the radiation field. The energy conservation equation accounts for flux-limited electron heat conduction and electron-ion energy relaxation (flux limiter $f = 0.06$). Radiation transfer is calculated using a flux-limited multi-group diffusion approximation with 20 frequency groups.  The atomic physics code IONMIX~\cite{MACFARLANE1989} was used to generate the opacity tables (Planck absorption and emission, and Rosseland group opacities), as well as the mean charge numbers, and equation-of-state tables for pressure and internal energy as functions of ion number density and plasma temperature.

The laser propagation is described in the geometrical optics approximation by a ray tracing routine, and the laser energy is deposited on the grid in the sub- to near-critical density regions according to the inverse Bremsstrahlung absorption. Generally, this method defines the beam and traces the laser rays in 3D geometry, and then projects the deposited energy back on the $r$-$z$ plane. In case of normal laser incidence, however, a pure 2D algorithm may be applied. Resonant absorption was not taken into account, as its contribution would be marginal for the  interaction conditions presented here.

\begin{figure}[!t]
\centerline{\includegraphics[width=0.95\textwidth]{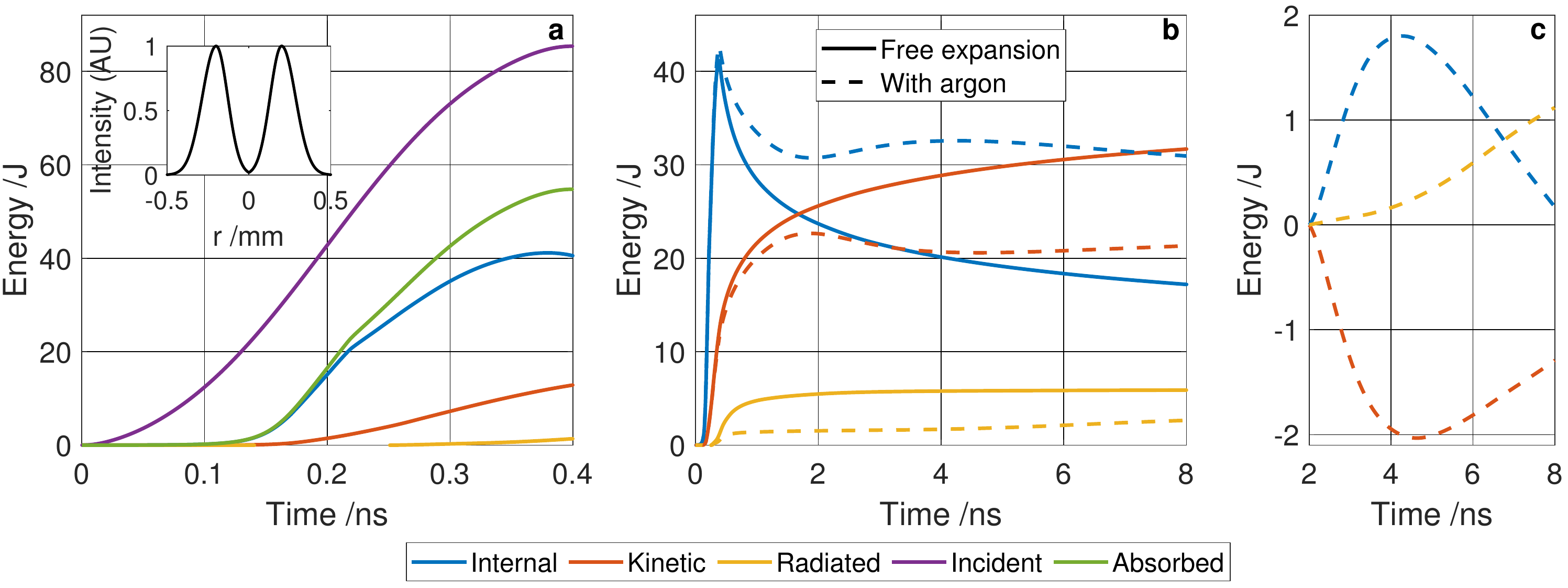}}
\caption{Simulation results showing the energy balance of the incident laser energy in the plasma. a) Energy balance for the free expansion of the copper plasma during the laser irradiation. The insert shows the radial profile of the laser intensity on the target. b) Energy balance with (dashed lines), and without the secondary argon target (solid lines). The initial interaction of the jet with the background argon plasma occurs around $t = 2$\,ns. c) The evolution of the energy balance following the shock formation. The curves have been shifted to have their zero value at $t = 2$\,ns in order to better illustrate the variation in energy during the shock evolution. }  \label{SimE}
\end{figure}

The simulation setup consisted of a 200\,$\mu$m thick copper foil, and a 3\,mm thick argon layer of constant density, separated by a 1.75\,mm gap. The argon density of $n_{\rm Ar} = 2\times10^{18}$\,cm$^{-3}$ used in the simulations, corresponds to a pressure of $p = 20$\,bar in the experiment. An energy of $E_{\rm L} = 85$\,J, wavelength $\lambda = 0.438\,\mu$m, and normal incidence on target with an f-number of 3, were chosen for the laser beam in accordance with the described experiment. The laser pulse had a Gaussian temporal profile with a full width at half maximum (FWHM) of 250\,ps, and truncated wings, giving a full duration of 400\,ps. Similar to the experiment, the laser had an annular intensity distribution on target~\cite{Kmetik2012}, with a radius of $r_{\rm L} = 300$\,$\mu$m (FWHM), as shown in the insert in figure~\ref{SimE}a. The duration of the simulation extended from the initial laser-target interaction, to the subsequent shock formation and evolution over 20\,ns. 

The time-dependence of the laser absorption, as well as the evolution of the internal, kinetic, and radiated energies, are shown in figure~\ref{SimE}. During the laser irradiation, as shown in figure~\ref{SimE}a for the free expansion of copper, approximately two thirds of the incident energy is absorbed. This initially goes into plasma heating, with the internal energy reaching $E_{\rm int} \sim 40$\,J at the end of the laser pulse. Later on, this is converted into the kinetic energy of the expanding plasma, eventually exceeding 30\,J. Similarly, only a small fraction of the energy goes into the X-ray emission during the laser pulse, but increases rapidly during later plasma expansion, attaining $E_{\rm rad} \sim 6$\,J.  

The time-evolution of the energy balance over 8\,ns is given in figure~\ref{SimE}b, with (dashed lines), and without (solid lines) the gas target implemented in the code. The gas is initially ionized by X-ray radiation emitted by the hot copper plasma during the laser pulse, heating it to a temperature of a few electronvolts. Eventually, the electron density increases sufficiently for the argon to be further ionized through direct collisional absorption of the incident laser light. The background argon plasma thereby reaches electron temperatures of several hundreds of electronvolts. 

The interaction of the copper jet with the argon plasma starts at around $t = 2$\,ns in figure~\ref{SimE}b. The shock formation is seen as a slight decrease of the plasma kinetic energy, and a significant increase in the internal energy, compared to the case without argon. Radiation losses are also significantly smaller, with approximately 2\,J going going into X-ray emission. Figure~\ref{SimE}c shows the energy balance during the subsequent shock evolution. Here, the zero energy value of the curves have been shifted to coincide with this time point, in order to better illustrate the energy exchange. The energy variation of approximately 2\,J during the time interval from 2 to 4 ns, shows the transfer of the kinetic energy of the jet into the internal energy of the shock. Moreover, the increase of 1\,J in the radiated energy over 6\,ns indicates the importance of radiation losses in the process of the shock formation and evolution.  

As described in the following sections, the observed jet and shock structures, and their plasma parameters, indicate that radiative effects dominate in the simulations. This is not the case for the experiment, and as a result, the electron temperatures and densities obtained in the simulations are significantly higher in the argon plasma, and lower in the copper. This is likely due to the opacity tables applied in the simulations resulting in an overestimation of radiative transfer compared with the experiment.   
For the experiment, the extent to which radiative processes would be expected to influence the dynamics, can be evaluated using the cooling parameter, given by the ratio of the hydrodynamic and cooling timescales. The hydrodynamic time, $t_{\rm h} = L/v$, is obtained from the characteristic length scale and velocity, and for a jet radius of $r_{\rm J} = 0.15$\,mm, and $v_{\rm J} = 700$\,km\,s$^{-1}$, it is $t_{\rm h} = 0.2$\,ns. Similarly, a shocked copper velocity of $v = 500$\,km\,s$^{-1}$ and working surface thickness of $L_{\rm ws} \leq 1$\,mm correspond to $t_{\rm h} \sim 2$\,ns. The radiation cooling time can be estimated by the ratio of the plasma thermal energy and the power of radiative losses, with the latter obtained from FLYCHK simulations. Line emission contributes significantly for the temperatures measured in the experiment, with the cooling time being two orders of magnitude longer if only bremsstrahlung is considered. For a jet with temperature $T_e = 100$\,eV, and density $n_e = 10^{19}$\,cm$^{-3}$, a radiation cooling time of $t_r \sim 1$\,ns is obtained, giving a cooling parameter of $t_h/t_r \sim0.2$. Similarly, shocked copper with $T_e = 200$\,eV and $n_e = 10^{20}$\,cm$^{-3}$, results in $t_r \sim 0.4$\,ns, and $t_h/t_r \sim 5$. This indicates that radiative effects may be important in the shocked copper in the experiment, but less so in the jet. The duration of the X-ray emission observed with the X-ray pinhole camera ($> 10$\,ns), also implies a radiation cooling time greater than, or on the order of, the hydrodyamic time scale in the experiment. Therefore, the dynamics of the jet and shock structure, formation, and evolution, cannot be solely attributed to either adiabatic or radiative effects, but likely arise from a combination of the two.

\subsection{Characterization of the copper jet}

\begin{figure}[!t]
\centering
\includegraphics[width=0.95\linewidth]{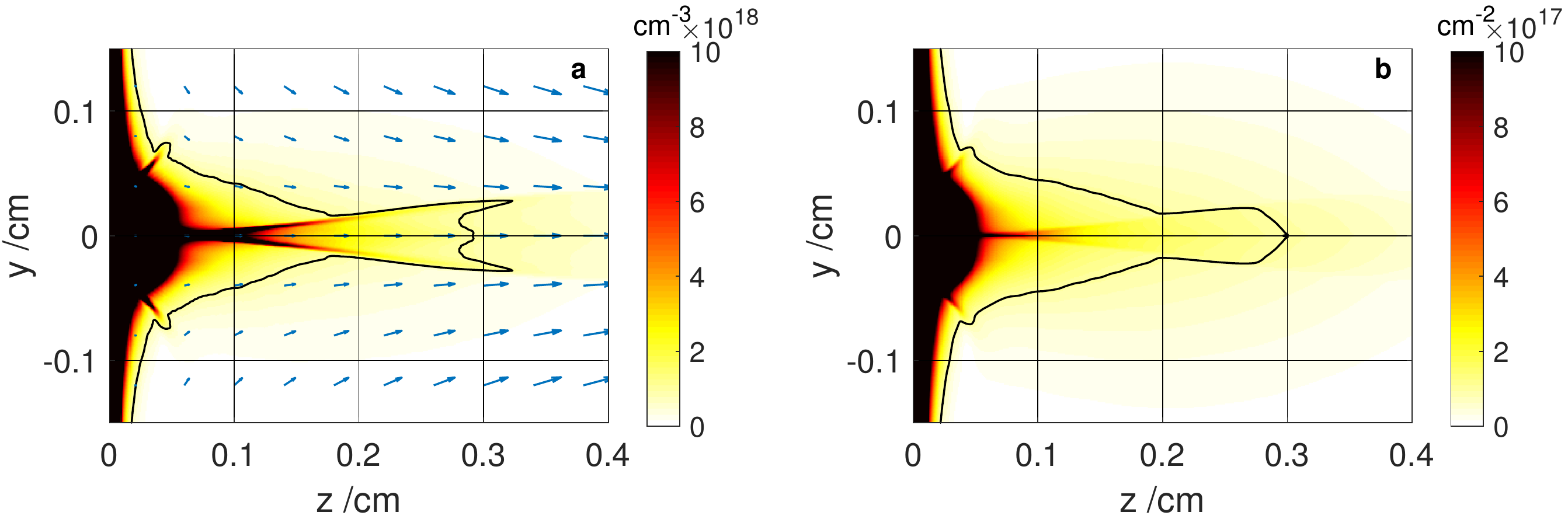}
\caption{Simulations of the free expansion of copper. a) Electron density, $n_e$, and b) line-integrated electron density, $N_e$, at $t= 5$\,ns. The black contour lines indicate $n_e = 10^{18}$\,cm$^{-3}$ in (a), and $N_e = 10^{17}$\,cm$^{-2}$ in (b). The blue arrows give the direction and relative amplitude of the flow velocity.}\label{Sim_Jet_ne}
\end{figure}

Before characterizing the shock formation and evolution in the simulations, it is useful to start with the simpler case of the jet, i.e. the free expansion of the copper plasma. The electron density of the jet at $t = 5$\,ns is shown in figure~\ref{Sim_Jet_ne}a, along with the corresponding line-integrated value in \ref{Sim_Jet_ne}b. The black contour lines indicate the electron density value $n_e = 10^{18}$\,cm$^{-3}$, and the line-integrated electron density value $N_e = 10^{17}$\,cm$^{-2}$, respectively. The blue arrows overlaid on the density plot in \ref{Sim_Jet_ne}a give the direction and relative amplitude of the flow velocity at these points. 

Similar to the experiment, the jet consists of a high-density region of ablated copper near the target, and a central collimated plasma flow extending along the target normal. The smaller jet-like structures on the sides of the plasma plume are also present in the simulation, though less pronounced than in the experiment. This is a typical feature arising from the collision of the radially expanding hot plasma with a colder plasma produced at the edges of the laser spot. 

The average jet velocity can be obtained in a similar way to the experiment. If the contour line in figure~\ref{Sim_Jet_ne}b is approximated as the tip of the jet, the position of $z = 0.3$\,cm at $t = 5$\,ns, yields an average jet velocity of $v_{\rm J} = 600$\,km\,s$^{-1}$. This is in good agreement with the velocity obtained in the experiment using the optical imaging diagnostic, and the X-ray streak camera. 

As can be seen in the quiver plot in figure~\ref{Sim_Jet_ne}a, the plasma motion also has a velocity component directed radially inwards. At the target, it due to the annular laser beam geometry, where the collision and compression of the converging copper plasma along the laser axis leads to the formation of the jet.  In addition, strong radiative cooling helps preserve the narrow shape of the jet over millimeter distances. However, recombination also leads to a decrease in the electron density at the center of the jet further from the target. Consequently, the initially narrow jet becomes broader, with a hollow cone-like structure at the edges. This is not observed in the experiment, and the lower charge, along with the order of magnitude lower copper temperatures obtained in the simulations, suggest it is due to differences in the radiation cooling time. Since this structure is less pronounced in the line-integrated density plot in figure~\ref{Sim_Jet_ne}b, another possibility is that it is due to the reduced spatial resolution for the density plots obtained with Abel inversion in previous jet experiments at PALS.

\subsection{Shock formation through jet-argon interaction}

\begin{figure}[!t]
\centering
\includegraphics[width=0.95\linewidth]{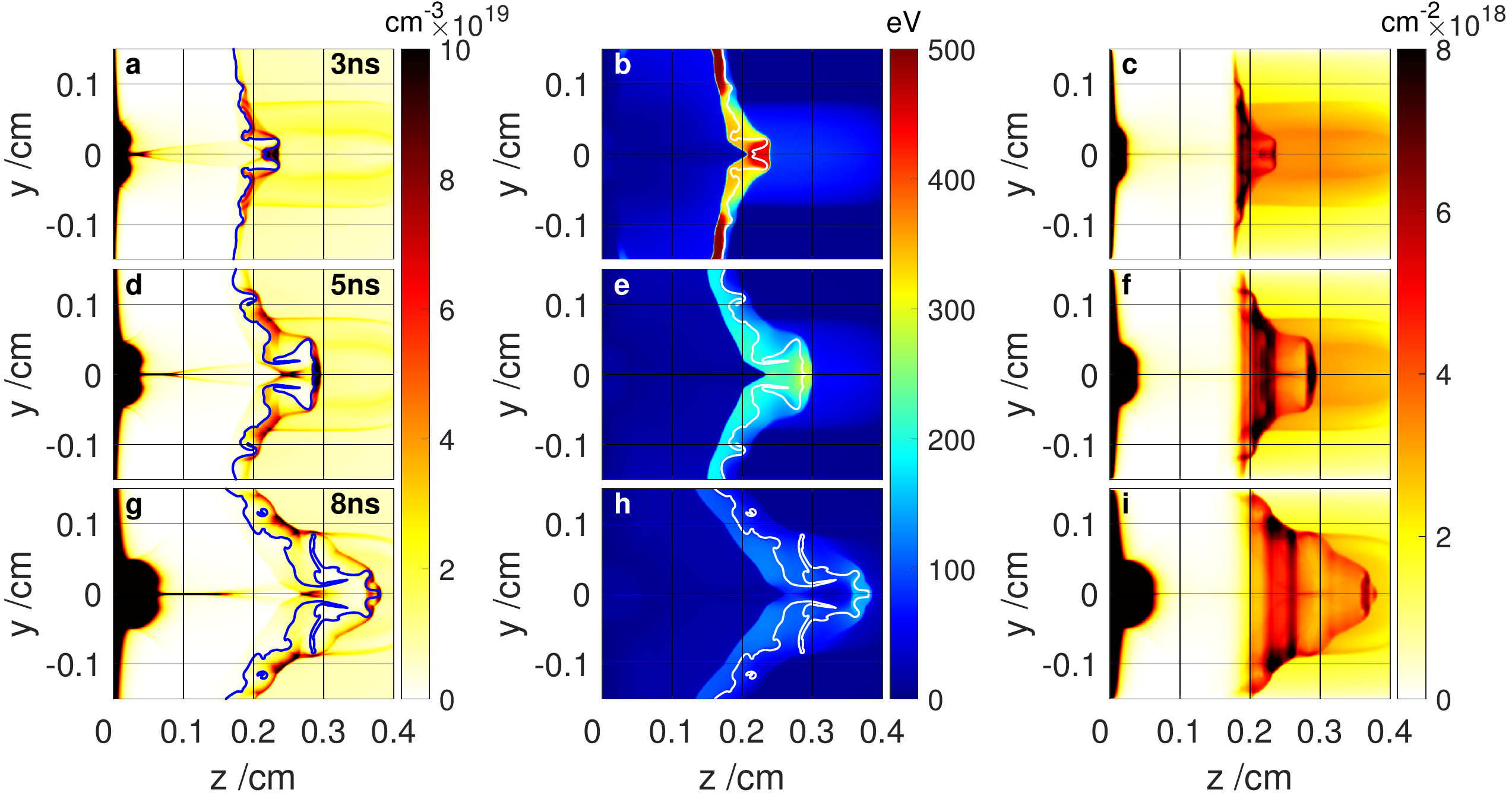}
\caption{Simulations of the shock formation and evolution. Electron density (left column), electron temperature (middle column), and line-integrated electron density (right column), at $t = 3$\,ns (top row), $t = 5$\,ns (middle row), $t = 8$\,ns (bottom row). The contour lines in the left and middle columns indicate the contact surface between copper and argon.}\label{Sim_Shock_ne}
\end{figure}

As in the experiment, the interaction of the copper jet with the argon gas target leads to the formation of a shock. The simulated electron density, line-integrated electron density, and electron temperature of the shock are shown for three time instances in figure~\ref{Sim_Shock_ne}. The contact surface between the copper and argon plasma is indicated by the contour lines in the electron density and temperature plots. 

Similar to the the experimental observations, the contribution of the laser heating to the ionization of the background argon is clearly visible in the simulation plots. Since the focal spot is located behind the target surface, additional heating is induced in the argon layer by the focused laser light reflected from the initially flat copper target. This, combined with the radial expansion of the background plasma due to the cooling effect, results in the visible decrease of the electron density at the center. It is not clear if these small-scale structures are also present in the experiment, however, the large-scale non-uniformity of the argon plasma is seen to influence the shock structure and evolution for both. In particular, the radial extent of the shock is influenced by the size and structure of the background plasma. 

The jet comes in contact with the argon plasma after approximately 2\,ns, and at $t = 3$\,ns, the bow shock has reached $z = 2.4$\,mm (top row in figure~\ref{Sim_Shock_ne}). Based on the contour line of the contact surface, a strongest shock compression occurs in a narrow region of argon, with the accumulation of shocked copper observed in the experiment being less pronounced in comparison. As the copper plasma slows down, and starts flowing back, a vortex is formed, and the shock front cools and spreads radially. The shape of the shock is also highly influenced by the jet structure, which in turn is affected by the interaction with the gas. As the jet traverses the argon plasma, a refocusing of the flow can be observed, leading to a local density increase of copper. The conical/arrow-like shape of the shocked copper at this location is similar to what is seen in the experiment (e.g. figure~\ref{Shock}). Since the non-uniform argon density profile of the gas target used in the experiment extends all the way to the target, the refocusing effect can also be expected to play a significant role in the jet formation. The improved jet collimation at low gas pressures can therefore likely be attributed to this. The line-integrated density plots in figure~\ref{Sim_Shock_ne} provide comparison with the overall shape of the shock seen in the experiment (e.g. figure~\ref{Shock}). In contrast to the experiment, the narrow argon bow shock has the greater contribution to the line integrated density, and hence the structure of the jet and shocked copper are not as easily discernible in these plots. However, the extent of the working surface of the combined argon and copper shock along the target normal is comparable to the experiment. 

\begin{figure}[!ht]
\centering
\includegraphics[width=0.95\linewidth]{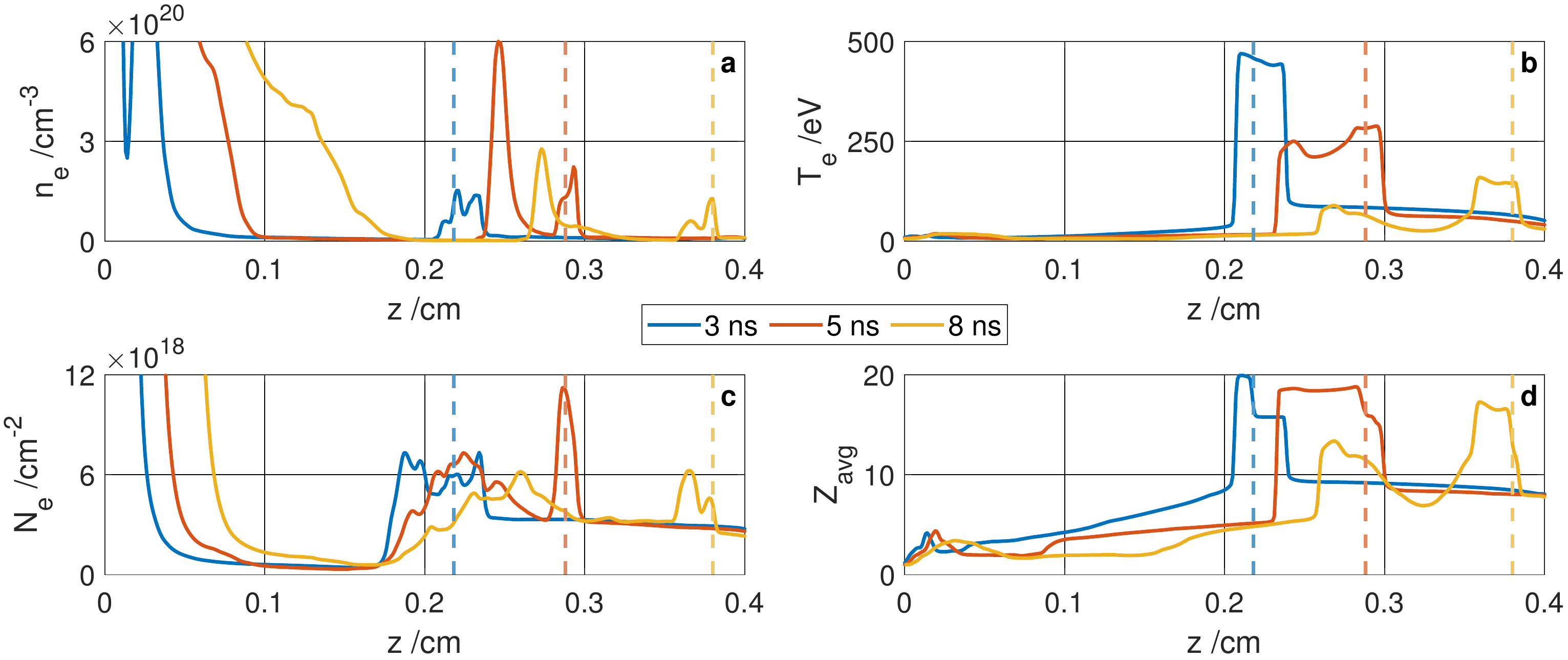}
\caption{Line-outs along the z-axis at y = 0 for: a) electron density, b) electron temperature, c) line integrated electron density, and d) mean ion charge, at $t =$ 3, 5, and 8\,ns. The vertical dotted lines indicate the location of the contact surface between copper and argon}\label{Sim_Shock_1D}
\end{figure}

The shock velocity can be obtained similarly to the experiment by tracking the location of the contact surface between frames. It decreases from $v \sim 350$\,km\,s$^{-1}$ at $t = 3$\,ns to $v \sim 230$\,km\,s$^{-1}$ at $t = 8$\,ns, which is in good agreement with the experimental values. The shock position, radial extent, and evolution are also comparable to the experimental observations. 
However, there are also some quantitative differences in temperature and density between the two, and in particular for the argon plasma, as shown by the line-outs along the target normal in figure~\ref{Sim_Shock_1D}. The density of the copper jet at $t = 3$\,ns goes from  $n_{e} \sim 10^{19}$\,cm$^{-3}$ at $z = 0.1$\,cm, to $n_{e} \sim 5\times10^{18}$\,cm$^{-3}$ at $z = 0.2$\,cm. This is slightly lower, but still within a factor of two of the jet density in the experiment. The overall copper temperature is also lower, with $T_e \sim 35$\,eV at $z = 0.2$\,cm, and the target having cooled to only a few electronvolts after 3\,ns. This is explained by the stronger radiation cooling and recombination effects in the simulations. As the jet progresses further into the argon, the refocusing of the copper leads to a local increase in density to $n_{e} \sim 6\times10^{20}$\,cm$^{-3}$ at $t = 5$\,ns. This can be attributed to additional ionization from the collision of copper plasma on axis, which is supported by the corresponding peak of the charge and electron temperature seen in this region. The electron temperatures at these later times are within the range observed in the shocked copper plasma at 3\,ns in the experiment.

The background argon electron density is on the order of $n_{e} \sim 10^{19}$\,cm$^{-3}$, and the electron temperature approximately $T_e \sim 80$\,eV, at $t = 3$\,ns. This is only a factor of two higher density than expected from the experiment, but at an order of magnitude higher temperature. The overestimation of the argon temperature and density are also seen in the shock. At $t= 3$\,ns the shocked argon density is on the order of $n_{e} \sim 10^{20}$\,cm$^{-3}$, which corresponds to a factor of ten increase compared to the ambient plasma. The argon temperature is approximately $T_e \sim 450$\,eV, and the argon charge $Z \sim 16$. In the simulations, these values arise from a combination of the shock compression, pressure ionization, and radiation cooling of the dense plasma. From the line-outs. it is also clear that the argon shock is narrower than in the experiment. At $t = 3$\,ns it has an extent of approximately 0.2\,mm, and decreasing further as the shock evolves. The weaker argon shock compression, along with the wider bow shock in the experiment, indicate that the shock heating and ionization of the gas are not as strong as in the simulations.

\begin{figure}[!b]
\centering
\includegraphics[width=0.99\linewidth]{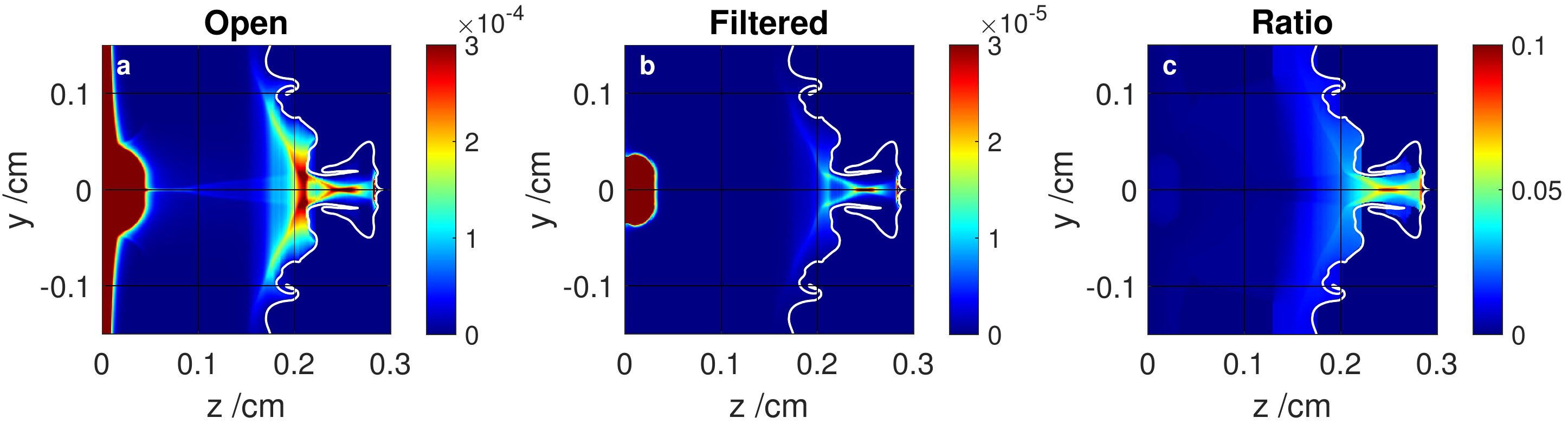}
\caption{Virtual X-ray pinhole diagnostic, showing the line-integrated copper emission at $t = 5$\,ns for the photon energy range 10 - 1000\,eV. a) Unfiltered emission; b) with a 0.2\,$\mu$m Fe filter; and c) their ratio. The values are normalised to the maximum of the unfiltered emission. The contour line indicates the contact surface between copper and argon}\label{Sim_Shock_Xray}
\end{figure}

The simulations can also be applied to generate a virtual X-ray diagnostic, which can be used for comparison with the X-ray pinhole images in the experiment (e.g. figure~\ref{PinholeShock}  and figure~\ref{Ratio2D}). Figure~\ref{Sim_Shock_Xray} shows the open and unfiltered line-integrated X-ray emission obtained from FLYCHK calculations of the emission spectra using the simulated electron temperatures and densities at $t = 5$\,ns in figure~\ref{Sim_Shock_ne}. As the copper emission dominates in the experiment, the emission from argon has been excluded in the line-integration. The overall shape, and position of the emitting zones are similar to the experiment, with the closest resemblance seen for the case with a 40\,bar backing pressure. The emission is strongest in the high density region near the target, with peaks in the shock region corresponding to the locations of local density maxima. Similar to the measurements at 40\,bar, the ratio is also greatest in shock. Though it is higher than in the experiment, this is due to differences in copper temperature and density, and the values are consistent with the ratio plot in figure~\ref{RatioComp}b.  

\section{Summary and conclusions}\label{SumSec}

This experiment was dedicated to the study of shocks formed through the interaction of laser-produced supersonic plasma jets with an ambient argon plasma, with particular emphasis on investigating their radiative properties. The jet and shock parameters were characterized using a combination of optical and X-ray diagnostics, allowing for a detailed study of their structure and evolution, from the initial ionization of the background plasma, and formation of the jet, to their subsequent interaction over a timescale on the order of 15\,ns. 

Interferometry, shadowgraphy, and X-ray measurements showed a strong shock compression in the copper plasma, with a factor of ten density increase compared to the argon background plasma. In comparison, the argon shock was weaker, with a density ratio of less than two. The shocked argon had an extent of $\sim 1$\,mm in the initial stages of the interaction, with the working surface narrowing as the faster moving copper plasma caught up to, and eventually broke through, the argon bow shock. 

The characteristic velocities of the jet and shock were obtained using an X-ray streak camera, and matched the propagation of the leading edge of the copper plasma observed in shadowgraphy. The shock formation was clearly seen in the streak, and the location and duration of the X-ray emission was consistent with X-ray pinhole imaging. A typical jet velocity of 400 - 700\,km\,s$^{-1}$ was observed, with a slower initial velocity seen in the presence of argon gas, and shock velocities around 300 - 500\,km\,s$^{-1}$. 

The radiative characteristics of the shock were studied in detail with X-ray pinhole imaging. The shock structure and evolution were found to be highly dependent on the initial neutral gas density, both through the interaction characteristics during the shock formation, as well as its influence on the initial jet formation and structure. The gradient of the non-uniform gas target extended all the way to the target, where the low density argon restricted radial expansion. This introduced a refocusing of the copper flow, yielding a narrower and better collimated jet for low gas pressures. For higher backing pressures, the gas also significantly affected the axial expansion, and the shock occurred closer to the target for the same time, with the dense plasma in the jet-forming region near the target cooling faster. Additionally, the ablated plasma at the periphery of the focal spot became sufficiently deviated towards the target normal to reach, and interact with, the high density region of the gas target, and thereby influence the shock structure.

Filtered X-ray pinhole imaging allowed for estimates of the electron temperature, which indicated $T_e = 50 - 150$\,eV in the jet, and $T_e \lesssim 150$\,eV in the shock-heated copper plasma. Comparison with interferometry showed that the X-ray emission mainly originated from the hot copper plasma, with little argon emission seen in the open pinhole frame in comparison, and none at all in the filtered one. This suggests a comparatively low electron temperature in the argon plasma, but could also be a result of re-absorption of emission in the colder ambient plasma in front of the diagnostic. Further study, implementing additional diagnostics, would be required to reliably determine the argon temperature. 

The formation and evolution of the jet and shock were simulated using the radiation-hydrodynamic code FLASH~\cite{Fryxell_2000, FLASH1, FLASH2}, thereby providing further insights into the physical processes involved. Comparison with the experimental observations also provided a validation of the code, as well as highlighting potential avenues for further development. A good qualitative agreement of the jet and shock characteristics was seen, with comparable velocities, and a shock evolution with similar radial extent and axial position. A refocusing of the copper flow by the argon plasma, similar to the experiment, was observed. A virtual X-ray pinhole diagnostic also provided good comparison between the two. However, quantitative differences were also seen in the argon plasma parameters such as the shock density, temperature, and thickness. These likely originate from differences in the initial conditions, and an overestimation of radiative transfer in the simulations. A more in-depth study of these discrepancies would be useful in future simulation work to further improve the agreement with the experiment.

The experimental results can be applied for a better understanding of the mechanisms governing the formation and evolution of jets and shocks associated with astrophysical phenomena such as Herbig-Haro (HH) objects. While the significant difference in plasma parameters and characteristic time and length scales means that the HH objects cannot be directly reproduced in the laboratory, they are governed by the same hydrodynamic behavior, and a scaling between the two is possible~\cite{Ryutov1999, Ryutov2000, Ryutov2001}. For the hydrodynamic model to be valid, the interactions need to be in the collisional regime, and in the experiment, the interaction between the copper jet and the background argon plasma is dominated by copper-electron collisions. For a jet with velocity $v_{\rm J} = 500$\,km\,s$^{-1}$, traversing a relatively cold argon plasma with $T_e = 10$\,eV, $Z = 2$, and $n_{e} = 5 \times 10^{18}$\,cm$^{-3}$, the mean free path is on the order of $\lambda_{Cu-e} \lesssim 200\,\mu$m. As this is smaller than the working surface thickness of the shock, the collisionality requirement is met in the experiment. The hydrodynamic scaling also requires thermal diffusion and viscosity to be negligible, which can be evaluated using the dimensionless Reynolds and Peclet numbers. For the parameters measured in the experiment, these are on the order of $Re \sim 10^7$, and $Pe \sim 200$, which are both sufficiently high for this approach to be valid.

The hydrodynamic scaling implies that the morphology of the two systems will evolve similarly, albeit at different time and length scales, if they have comparable Mach numbers, $M = v_{\rm J}/c_{\rm s}$. In the experiment, for a background argon plasma with electron temperature of $T_e = 10$\,eV, a speed of sound $c_s \sim 15$\,km/s is obtained. A supersonic jet velocity of $v_{\rm J} = 500$\,km\,s$^{-1}$ then results in a Mach number of $M \sim 35$. Similarly for copper, an electron temperature of $T_e = 100$\,eV gives $c_{\rm s} \sim 60$\,km\,s$^{-1}$, and $M \sim 8$. The Mach number can vary significantly between different types of HH-objects, and is usually in the range of $M \sim 5 - 40$. Since the characteristic length and time both scale with the velocity, comparing the size of two systems with similar Mach numbers, provides a corresponding scaling of their temporal evolution. The experimental timescale on the order of 10\,ns would then roughly correspond to hundreds of years for a HH-object, with the evolution of the shock and jet structures corresponding to tens of years. The overall lifetime of HH-objects is often on the order tens of thousands of years, but they also exhibit a dynamic behavior, with the small-scale structures observed to evolve over timescales of only a few years in some cases~\cite{Reipurth2002, Hartigan2001, Hartigan2011, Bally2002}. The experimental observations, such as the formation, structure, and evolution of the jet and shock, should therefore be relevant for understanding similar processes in Herbig-Haro objects.

\ack
This research was partially supported by the Czech Republic MSMT targeted support of Large Infrastructures, ELI Beamlines Project LQ1606 of the National Programme of Sustainability II. S.W. was supported by the project Advanced research using high intensity laser produced photons and particles (ADONIS) (CZ.02.1.01/0.0/0.0/16\_019/0000789), and by the project High Field Initiative (HiFI) (CZ.02.1.01/0.0/0.0/15\_003/0000449), both from the European Regional Development Fund. This research was also supported by the Access to the PALS RI under the EU LASERLAB IV (Grant Agreement No. 654148) and LASERLAB AISBL (Grant Agreement No. 871124) projects; by the Ministry of Science and Higher Education, Republic of Poland (Decision NR 5084/PALS/2020/0), by the Ministry of Education. This work has been partially carried out within the framework of the EUROfusion Consortium and has received funding from the Euratom research and training program 2019-2020 under grant agreement No 633053 (EUROfusion project ENR-IFE19.CEA-01). The views and opinions expressed herein do not necessarily reflect those of the European Commission.

\section*{References}
\bibliographystyle{unsrt}
\bibliography{refs}

\end{document}